\documentstyle[multicol,aps,prl,psfig]{revtex} 

\renewcommand{\narrowtext}{\begin{multicols}{2} \global\columnwidth20.5pc}
\renewcommand{\widetext}{\end{multicols} \global\columnwidth42.5pc}
\def\ep{\epsilon}

\def\th{\theta}

\def\ka{\kappa}
\def\la{\lambda}

\def\rh{\rho}

\def\ta{\tau}

\def\ph{\phi}

\def\om{\omega}
\def\Ga{\Gamma}
\def\De{\Delta}

\def\La{\Lambda}
\def\Si{\Sigma}

\def\Ps{\Psi}
\def\Om{\Omega}

\def\cl{{\cal L}}

\def\cF{{\cal E}}

\def\cV{{\cal V}}

\def\fr#1#2{{{#1} \over {#2}}}
\def\frac#1#2{\textstyle{{{#1} \over {#2}}}}

\def\prt{\partial}

\def\half{{\textstyle{1\over 2}}}
\def\lsim{\mathrel{\rlap{\lower4pt\hbox{\hskip1pt$\sim$}}
    \raise1pt\hbox{$<$}}}
\def\gsim{\mathrel{\rlap{\lower4pt\hbox{\hskip1pt$\sim$}}
    \raise1pt\hbox{$>$}}}

\def\Re{\hbox{Re}\,}
\def\Im{\hbox{Im}\,}

\newcommand{\beq}{\begin{equation}}
\newcommand{\eeq}{\end{equation}}
\newcommand{\bea}{\begin{eqnarray}}
\newcommand{\eea}{\end{eqnarray}}
\newcommand{\rf}[1]{(\ref{#1})}

\def\cF{{\cal F}}

\begin{document}

\title{Analogue Models for T and CPT Violation 
in Neutral-Meson Oscillations}    
\author{V.\ Alan Kosteleck\'y and \'Agnes Roberts}
\address{Physics Department, Indiana University, 
          Bloomington, IN 47405, U.S.A.}
\date{IUHET 428, December 2000} 
\maketitle

\begin{abstract}
Analogue models for CP violation
in neutral-meson systems are studied in a general framework.
No-go results are obtained for models in classical mechanics 
that are nondissipative or that involve one-dimensional oscillators. 
A complete emulation is shown to be possible for
a two-dimensional oscillator with rheonomic constraints,
and an explicit example with spontaneous T and CPT violation 
is presented.
The results have implications 
for analogue models with electrical circuits.

\end{abstract}


\narrowtext

\section{Introduction}

Many quantum systems display oscillatory behavior.
Among the most interesting are the neutral mesons,
where oscillations between particles and antiparticles
can violate the product CP of charge conjugation (C)
and parity (P) symmetries.
In the $K$ system,
a small CP violation is experimentally seen 
\cite{ccft}.
It is associated with breaking of time-reversal symmetry T,
with the product CPT being preserved
\cite{sachs}.
In fact,
a complete formulation of CP-violating oscillations 
in the $K$ system allows
also CPT violation with T being preserved
\cite{lw,fn1}. 
A similar formulation can be developed for any of the neutral mesons
$K$, $D$, $B_d$, and $B_s$
\cite{kp2}.
This more general situation is of interest,
for example,
in the context of possible experimental signals from string theory
\cite{kp}.

Acquiring physical insight 
into the behavior of meson oscillations
in the presence of T and CPT violation is worthwhile.
One approach is to construct a simple analogue model
in classical mechanics that displays the key features 
of meson oscillations.
\it A priori, \rm
it seems most natural to adopt an intuitive picture 
based on an analogue model
in which the meson and its antimeson are represented
by two one-dimensional oscillators
interacting through some weak coupling. 
Indeed,
basic features of the CP-preserving physics
can correctly be modeled in this way
\cite{bw}.
However,
modeling T violation is more subtle 
\cite{jr,rs}.

In this work,
we investigate the issue of emulating both T and CPT violation
in neutral-meson systems 
via models in classical mechanics involving small oscillations.
We obtain several no-go results,
showing that complete emulation of the effective hamiltonian
describing the time evolution of a neutral meson is impossible 
using models with no damping
or using models involving two one-dimensional oscillators
with a large class of couplings.
This confirms and extends earlier results of Rosner
\cite{jr}.
In contrast,
models involving two-dimensional oscillations 
with appropriate constraints can 
display effects emulating both T and CPT violation simultaneously.
We give an explicit example
in which the violation arises spontaneously.

In the next section,
we present a few basic results 
needed for the subsequent analysis.
Section III discusses some no-go results.
The issue of spontaneous symmetry breaking
is considered in section IV.
The general analysis leading to a complete
emulation of neutral-meson effective hamiltonians,
including an explicit model,
is given in section V.
Section VI summarizes the results 
and discusses some open issues,
including the implications of our results 
for the challenge of emulating CP violation
with electric circuits.

\section{Basics}

In this section,
we introduce some basic results 
needed for the analysis in later sections. 
Following a discussion of relevant features of
the neutral-meson systems,
a few considerations appropriate 
for classical analogue models are presented. 

The four relevant neutral-meson systems are
$K^0$, $D^0$, $B^0_d$, and $B^0_s$.
In what follows,
we denote by $P^0$ the strong-interaction eigenstate
associated with any one of these. 
A general neutral-meson state is a linear combination of
meson and antimeson wave functions.
It can be represented as a two-component object $\Ps$,
with time evolution determined 
by a 2$\times$2 effective hamiltonian $\La$
according to 
$i \prt_t \Ps = \La \Ps$.
The eigenvectors $P_S$ and $P_L$ of $\La$ 
are the physical propagating states.
The hamiltonian $\La$
is composed of a hermitian mass matrix $M$ 
and a hermitian decay matrix $\Ga$:
$\La = M - \frac 12 i\Ga$.
Flavor oscillations and T violation are governed 
by the off-diagonal elements of $\La$,
while CPT violation is controlled by the difference 
between its diagonal elements.

A widely used parametrization of $\La$ is
\cite{lw}
\beq
\La = \left(
\begin{array}{lr}
- iD + E_3 & E_1 - iE_2 \\ & \\
E_1 + iE_2 & -iD - E_3
\end{array}
\right),
\label{laparm}
\eeq
where $D$, $E_1$, $E_2$, $E_3$ are complex.
In this parametrization,
T violation occurs when
$(E_1E_2^* - E_1^*E_2) \neq 0$
while CPT violation occurs when $E_3 \neq 0$.
In terms of real and imaginary components,
T violation occurs when
\beq
(\Re E_2\Im E_1 - \Re E_1\Im E_2)\neq 0,
\label{tviol}
\eeq
and CPT violation when 
either or both of \beq
\Re E_3 \neq 0,\qquad \Im E_3 \neq 0,
\label{cptviol}
\eeq
is satisfied.
There are therefore three independent real quantities
determining CP violation in neutral-meson systems. 

The basic goal in achieving the construction 
of a suitable analogue model
is to obtain an oscillating system in classical mechanics
with a characteristic matrix 
reproducing the features of the effective hamiltonian $\La$.
It is therefore useful to consider 
the extent to which the form \rf{laparm} of $\La$
can be modified without affecting the underlying meson physics.

One flexibility in the form of $\La$ arises because the  
$P^0$ and $\overline{P^0}$ wave functions 
are eigenstates of the strong interactions,
which preserve strangeness, charm, and beauty.
For a given system,
the phases of the two wave functions can be rotated
by equal and opposite amounts without observable consequences.
This rotation induces a corresponding change in the
phase of the off-diagonal components of $\La$,
which acts to mix $E_1$ and $E_2$
but preserves the combination
$(E_1E_2^* - E_1^*E_2)$
measuring T violation.
It follows that a satisfactory analogue model
needs to contain features corresponding to this
phase-independent measure of T violation.
Note, however,
that the phase rotation cannot mix components 
of the mass matrix $M$ with
components of the decay matrix $\Ga$.

Another flexibility relevant to the construction of an analogue model
is the choice of basis for the meson system.
The effective hamiltonian $\La$ in Eq.\ \rf{laparm}
is given in the $P^0$-$\overline{P^0}$ basis.
However,
other unitarily equivalent bases can also be chosen.
A basis transformation by a unitary matrix $U$
converts the state $\Ps$ into 
$\Ps^\prime = U \Ps$ and results in an effective hamiltonian
$\La^\prime = U^\dagger \La U$.
Appropriate choices for $U$ can modify the location of  
the parameters for T and CPT violation in the effective hamiltonian,
which may have some advantages in matching to an analogue model.
Note that,
like the phase-rotation freedom,
the transformation by $U$ cannot mix components 
of the mass matrix $M$ with
components of the decay matrix $\Ga$.

As an example, 
consider a CP-eigenstate basis $P_1$, $P_2$
obtained from $P^0$, $\overline{P^0}$ 
via the unitary transformation
\beq
U = 
\fr 1 {\sqrt{2}}
\left( 
\begin{array}{lr}
1& 1 
\\ & \\
1 & -1 
\end{array}
\right).
\label{umatrix}
\eeq
In the new basis, 
the effective hamiltonian $\tilde\La$ becomes
\beq
\tilde\La = \left(
\begin{array}{lr}
- iD + E_1 & E_3 + iE_2 \\ & \\
E_3 - iE_2 & -iD - E_1
\end{array}
\right).
\label{tilaparm}
\eeq
The elements of $\tilde\La$ are similar to those of $\La$,
except that $E_1$ and $E_3$ have been interchanged.
This conversion to $\tilde\La$
was used by Rosner and Slezak
\cite{rs}
to show that a modified damped Foucault pendulum 
can be identified as an analogue model 
for T violation in the CP-eigenstate basis of a meson system.

Other choices can be made.
For example,
combining the choice of CP-eigenstate basis with
a phase rotation by $\exp(i\pi/4)$ of the $P^0$ wave function 
and an opposite rotation of the $\overline{P^0}$ wave function
yields an effective hamiltonian $\hat\La$ given by
\beq
\hat\La = \left(
\begin{array}{lr}
- iD + E_2 & E_3 - iE_1 \\ & \\
E_3 + iE_1 & -iD - E_2
\end{array}
\right).
\label{hatlaparm}
\eeq
This corresponds to a modification of $\La$ involving
a cyclic permutation of the three parameters:
$(E_1,E_2,E_3) \to (E_3,E_1,E_2)$.

In developing an analogue model,
we adopt the notion that the 
behavior of the strong-interaction eigenstates 
$P^0$, $\overline{P^0}$ 
can be modeled classically
by identifying them with harmonic oscillators
in two generalized coordinates.
The energies of the meson eigenstates
are emulated by the oscillator frequencies,
while the meson decay rates are paralleled by the oscillator dampings. 

Since the strong interactions preserve CPT,
in the absence of CP violation
the two frequencies and decay rates are expected to 
be equal. 
The idea is to model the presence of CP violation
by introducing appropriate couplings 
between the two classical oscillators.
Following ideas concerning CPT violation in the context
of conventional quantum field theory
\cite{kp2,ck},
we regard it as desirable 
to obtain CPT violation spontaneously in an analogue model.
In fact,
we show below that it is also possible 
to generate T violation spontaneously.

We limit attention to classical models involving
small oscillations about equilibrium
with linear equations of motion.
Assuming harmonic behavior, 
the linear generalized coordinates $q_1$, $q_2$ 
can be combined in two-component form 
as $Q=\Re[A\exp(i\om t)]$,
where $A$ is a complex two-component object.
The equations of motion can then be expressed by the
action of a 2$\times$2 matrix $X(\om)$ on $A$,
as $XA=0$.
The matrix $X$ is the characteristic matrix of the 
classical oscillator.

A suitable analogue model for a neutral-meson system
is one for which the characteristic matrix $X$ reproduces
the features of the meson effective hamiltonian.
In comparisons between the analogue model and the meson system,
it is useful to adopt 
a form for $X$ analogous to that for $\La$ in Eq.\ \rf{laparm}.
We therefore introduce the parametrization 
\beq
X = \left(
\begin{array}{lr}
- iA + B_3 & B_1 - iB_2 \\ & \\
B_1 + iB_2 & -iA - B_3
\end{array}
\right),
\label{xparm}
\eeq
where $A$, $B_1$, $B_2$, $B_3$ are complex.

The reader is cautioned that, 
despite the similarity 
of the parametrizations \rf{laparm} and \rf{xparm},
the detailed physical meanings of $\La$ and $X$ differ.
For instance,
$\La$ involves a first-order time development
while $X$ involves a second-order one.
A related point is that
the meson state $\Ps$ is intrinsically complex,
with physical observables being related to the
norm of the probability amplitude.
In contrast,
the mechanical coordinate $Q$ is real,
and the corresponding amplitude $A$ is complex only
as a convenient artifact.
For example,
opposite phase rotations between the two coordinates
could produce a physically inequivalent result
in the classical analogue model,
whereas similar phase rotations on the strong-interaction eigenstates 
have no physical effect in the meson system.

\section{Models without damping}

The intrinsic physical differences between the quantum system
and the classical model
might seem sufficiently severe to exclude emulation of
subtle effects such as T and CPT violation.
Indeed,
several no-go results can be obtained
concerning the existence
of an acceptable analogue model for $\La$ 
under various circumstances.
In this section,
we discuss obstacles to the development of an analogue model
in the absence of damping forces.
The effects of dissipation are considered in section V.

Consider first a lagrangian $\cl$ describing
small linear oscillations in a conservative classical-mechanical system.
For present purposes,
we restrict attention to a system with two degrees of freedom,
although some of our formalism and
results apply more generally.

Linearity implies that $\cl$ is quadratic in 
the real generalized coordinates $Q(t)$
and the first time derivatives $\dot Q \equiv dQ/dt$.
It can be written as
\beq
\cl = \half \dot Q ^T T \dot Q
+ \half \dot Q^T G Q
- \half Q^T V Q,
\label{lagr}
\eeq
where $T$, $G$, and $V$ are square matrices 
of the same dimension as $Q$.
By inspection,
$T$ and $V$ are symmetric, 
while $G$ is antisymmetric.
Since $\cl$ is real,
all three matrices can be taken real without loss of generality.
We call $T$, $G$, and $V$ the
kinetic, gyroscopic, and potential matrices,
respectively.
Note that $G$ violates classical time-reversal symmetry.

The Euler-Lagrange equations of motion obtained from 
the lagrangian $\cl$ are
\beq
T \ddot Q + G \dot Q + VQ=0.
\label{elnodamp}
\eeq
The gyroscopic matrix $G$ does \it not \rm represent damping,
despite its association with $\dot Q$,
because it is derived from a lagrangian
and the corresponding generalized force is conservative.
For harmonic solutions with $Q = \Re [A \exp(i \om t)]$
Eq.\ \rf{elnodamp} becomes $XA = 0$,
where the characteristic matrix $X$ has the form
\beq
X = - \om^2 T + i\om G + V.
\eeq
This matrix is hermitian and so can be diagonalized
with real eigenvalues.
The normal-mode frequencies are obtained from the condition
$\det X(\om) = 0$,
which is a quadratic equation in $\om^2$.
The absence of damping physically implies that 
there are two real normal-mode frequencies,
and this can be confirmed by inspection of the discriminant
of the general solution for $\om^2$. 

In terms of the parametrization \rf{xparm} of $X$,
we find:
\bea
\Re A &=& \Im B_1 = \Im B_2 = \Im B_3 = 0,
\nonumber\\
\Im A &=& -\half \om^2 (T_{11} + T_{22}) + \half (V_{11}+ V_{22}),
\nonumber\\
\Re B_3 &=& -\half \om^2 (T_{11} - T_{22}) + \half (V_{11}- V_{22}),
\nonumber\\
\Re B_1 &=& - \om^2 T_{12} + V_{12},
\nonumber\\
\Re B_2 &=& - \om G_{12}.
\label{xnodamp}
\eea
The form of Eq.\ \rf{xnodamp}
permits several conclusions about the feasibility
of constructing nondissipative analogue models 
for CP violation in neutral-meson systems.
Next,
we discuss these conclusions for $T$ and CPT violation in turn.

To begin,
observe that Eq.\ \rf{xnodamp} includes
the result $\Im B_1 = \Im B_2 = 0$
for all possible conservative classical systems.
In contrast,
Eq.\ \rf{tviol} implies
that at least one of $\Im E_1$ and $\Im E_2$ must
be nonzero for T violation in a meson system.
It follows by comparison of Eqs.\ \rf{laparm}
and \rf{xparm}
that T violation in the $P^0$-$\overline{P^0}$ basis
with the effective hamiltonian $\La$ 
cannot be emulated by any nondissipative classical model
with two degrees of freedom.

Transformation to some other basis for the meson wave functions
offers more flexibility but remains insufficient.
For example,
in the CP-eigenstate basis 
a nonvanishing component $\Re E_1$
in the effective hamiltonian \rf{tilaparm}
can be modeled with $\Re B_3$,
but no means to model 
$\Im E_1$, $\Im E_2$, $\Im E_3$ exists.
The point is that neither the phase-rotation flexibility 
nor the choice of wave-function basis
can mix contributions to the mass matrix $M$ 
with those to the decay matrix $\Ga$,
as discussed in the previous section.
Since $\Im E_1$, $\Im E_2$, $\Im E_3$ are contained in $\Ga$
while $\Re E_1$, $\Re E_2$, $\Re E_3$ are contained in $M$,
there is no means to convert one type of contribution to another.

We conclude that
\it 
it is impossible to emulate T violation 
in a neutral-meson system
with any nondissipative classical model
having two degrees of freedom.
\rm
In essence,
a successful analogue model for T violation in a meson system
must involve dissipation
because T violation in the meson system itself 
intrinsically involves dissipative oscillations.

The situation for CPT violation has both similarities 
and differences.
Comparison of Eqs.\ \rf{laparm} and \rf{xparm}
shows that nonzero CPT violation in a meson system
involving $\Re E_3$ can be
emulated by a classical oscillator model
for which $\Re B_3 \neq 0$.
The result \rf{xnodamp}
reveals that it suffices to have a difference between the diagonal
elements of either the kinetic or the potential matrix.
This is straightforward to achieve
in a physical system.
In contrast,
an argument similar to that for T violation
demonstrates that 
\it
it is impossible to emulate
CPT violation involving $\Im E_3$ in a meson system 
with any nondissipative classical model
having two degrees of freedom.
\rm
This can again be traced to the association of $\Im E_3$
with the decay matrix $\Ga$
and hence with dissipation in the meson system.

The strength of these no-go results 
suffices to show the need for dissipative classical oscillations.
However,
before turning to issues pertaining to spontaneous breaking
and dissipation,
we present some remarks about gyroscopic terms
in the context of conservative systems.

For a completely general emulation of neutral-mesons systems, 
we deem it desirable to construct an analogue model
for which all eight real parameters in Eq.\ \rf{xparm}
are nonzero.
The result \rf{xnodamp} shows that 
in the absence of damping
a nonzero gyroscopic matrix $G$ 
is needed to obtain a nontrivial $\Re B_2$.
In fact, 
this also holds in the presence of dissipation,
as is shown in section V.
Models without gyroscopic terms are therefore of lesser interest.
However,
gyroscopic terms appear in only a restricted class of models.
In particular,
there is no simple means of generating a nonzero $G$
in models involving two coupled one-dimensional oscillators,
as we discuss next.

Prior to linearization,
a general lagrangian describing two coupled one-dimensional oscillators
involves a kinetic term for each oscillator
and an interaction potential.
No gyroscopic term is present.
By definition,
the kinetic energy of a one-dimensional oscillator 
involves only one generalized coordinate,
so linearization of the kinetic pieces cannot generate the
cross-coupling needed for a nonzero $G$.
The potential term would therefore need to be the source of $G$. 
However,
the gyroscopic term is linear in the generalized velocity,
so any appropriate potential term must be velocity dependent.
This leaves only a restricted class of possibilities. 

It can be shown that $G$ makes 
no contribution to the hamiltonian,
so any acceptable velocity-dependent potential must describe forces
that do no work.
Forces that do no work and are described by a velocity-dependent potential
certainly exist.
A standard example is the Lorentz force on a charged particle moving 
in a magnetic field.
One might, 
for example,
consider a model involving two charged magnetic dipoles,
each restricted to move along a one-dimensional curve 
so that the only possible oscillations are indeed one-dimensional. 
The force $\vec F_{21} \propto \vec v_1 \times \vec B_2$ 
on one dipole is determined by its velocity $\vec v_1$ 
and by the field $\vec B_2$ of the other dipole,
as needed.
However,
this fails to generate directly a nonzero $G$ because   
$\vec F_{21} \cdot \vec v_1 \equiv 0$,
so the force is orthogonal to the oscillation. 

In short,
we find that 
\it
it is difficult and perhaps impossible to emulate
all eight parameters for a neutral-meson effective hamiltonian
with any classical model involving 
two coupled one-dimensional oscillators.
\rm
We conjecture that an impossibility proof could be constructed
on the basis that $G$ violates classical time-reversal invariance,
which imposes severe constraints on one-dimensional systems.
In any event,
it would be interesting to obtain an impossibility proof
or to provide a simple counterexample.

The above result provides strong motivation to turn instead
to an analogue model involving one two-dimensional oscillator.
In this case,
it is possible to generate a nonzero $G$ under suitable circumstances.

Before linearization,
the kinetic term of a two-dimensional model
typically involves both generalized coordinates.
If the equilibrium coordinates and configuration
are independent of time (scleronomous constraints)
and if there are no ignorable coordinates,
then the kinetic term is quadratic in generalized velocities
\cite{etw}
and so no gyroscopic term emerges upon linearization.
However,
for the special class of models 
with time-dependent (rheonomic) constraints,
linearization of the kinetic term can generate a nonzero $G$ matrix.
For example,
suppose the model involves small oscillations about a uniform motion,
characterized by a constant $v_0$ 
with dimensions of velocity.
Then, 
linearization of the terms quadratic in generalized velocities
can lead to expressions involving the product of $v_0$ 
and the oscillation velocity $\dot Q$.
These are linear in $\dot Q$ and under suitable circumstances
can yield a nonzero $G$.
Indeed,
the term `gyroscopic' refers to 
the appearance of a nonzero $G$ matrix 
in the description of small oscillations of uniformly rotating bodies.
Note that the classical T violation necessary for gyroscopic terms
emerges here as a result of the uniform motion.

\section{Spontaneous symmetry breaking}

In this section,
we discuss the issue of generating 
T and CPT violation spontaneously in the classical model.
The idea is to seek an analogue model with 
an initial configuration displaying no T or CPT violation,
but with a perturbative instability causing a natural
dynamical evolution to a stable configuration
in which small oscillations violate both T and CPT.
This parallels the mechanism for spontaneous breaking of CPT 
in string field theory 
\cite{kp}.
Following the discussion in the previous section,
we primarily restrict attention 
to the case of one two-dimensional oscillator
without appreciable dissipation.
The situation for viscous damping is considered 
in the next section.

Consider first a particle moving under the influence of gravity
on the interior of a spherical bowl 
that rotates with constant angular speed $\Om$
about the vertical axis.
The configuration with particle initially at the bottom of the bowl
is a solution to the equations of motion.
However,
an otherwise negligible friction between the particle and the bowl
makes this configuration perturbatively unstable 
if $\Om^2 > g/a$,
where $g$ is the gravitational acceleration 
and $a$ is the bowl radius. 
The position of stable equilibrium 
lies instead on the surface 
at a vertical distance $g/\Om^2$ 
below the center of the bowl.
This example,
introduced by H.\ Lamb in his paper on kinetic stability in 1908
\cite{hl},
provides a classical implementation of spontaneous breaking
of rotational symmetry.

By itself,
this example is unsatisfactory as the basis 
for an analogue model for CP violation in neutral-meson systems
because no restoring force is associated with 
a small horizontal displacement from the equilibrium position 
on the bowl's surface.
However,
a more general surface with noncircular horizontal cross section
can avoid this difficulty.
One might, 
for example,
consider a surface that is a spherical bowl at the bottom 
but that smoothly deforms into a surface
of uniform elliptical cross section as the height increases.
In this case only two equilibrium points occur,
located on the semi-major axis of the elliptical cross section.
In equilibrium,
the particle rotates with the bowl.
Small oscillations about either equilibrium point are stable
in both vertical and horizontal directions.

For an explicit analysis in the case of a suitable general surface,
we adopt cylindrical coordinates $(\rh,\ph,z)$
with origin at the bottom of the bowl.
Let the bowl's surface be determined by the equation
$f(\rh,\ph,z) = 0$,
where by assumption $f$ satisfies all the necessary convexity 
and smoothness conditions.
Then,
the motion on the surface of a particle of mass $m$ under gravity 
is determined by the lagrangian
\beq
\cl = \half m [\dot\rh^2 + \rh^2 ( \dot\ph + \Om)^2 + \dot z^2]
- m g z
+ m \la f(\rh,\ph,z),
\eeq
where $\la$ is a Lagrange multiplier.
See Figure 1.

For sufficiently large $\Om$,
spontaneous symmetry breaking occurs.
The equilibrium point $(\rh_0,\ph_0,z_0)$ is determined by  
the equations $f = 0$, $f_\ph = 0$,
and $\rh_0 \Om^2 + g f_{\rh}/f_{z}=0$
evaluated at the equilibrium point,
where subscripts on $f$ indicate partial derivatives.
Taking $\rh$ and $\rh_0\ph$ as generalized coordinates
for small oscillations
of frequency $\om$ about the equilibrium point,
a short calculation shows that 
the characteristic matrix has components 
\bea
X_{11} &=& 
- (1 + \Ga^2) \om^2 - \Om^2
\nonumber \\
&&\quad
- \la_0 (f_{\rh\rh} + 2 \Ga f_{\rh z} + \Ga^2 f_{zz}),
\nonumber\\
X_{12} = X_{21}^* &=&
- 2i\Om\om - \la_0(f_{\rh\ph} + \Ga f_{\ph z})/\rh_0,
\nonumber\\
X_{22} &=&
- \om^2 - \la_0 f_{\ph\ph}/\rh_0^2,
\label{xcomps}
\eea
where $\Ga = \rh_0\Om^2/g$, $\la_0 = g/f_z$,
and the partial derivatives are again evaluated 
at the equilibrium point.
Note the appearance of the off-diagonal gyroscopic terms $\pm 2i\Om\om$,
as expected.

\begin{figure}
\centerline{\psfig{figure=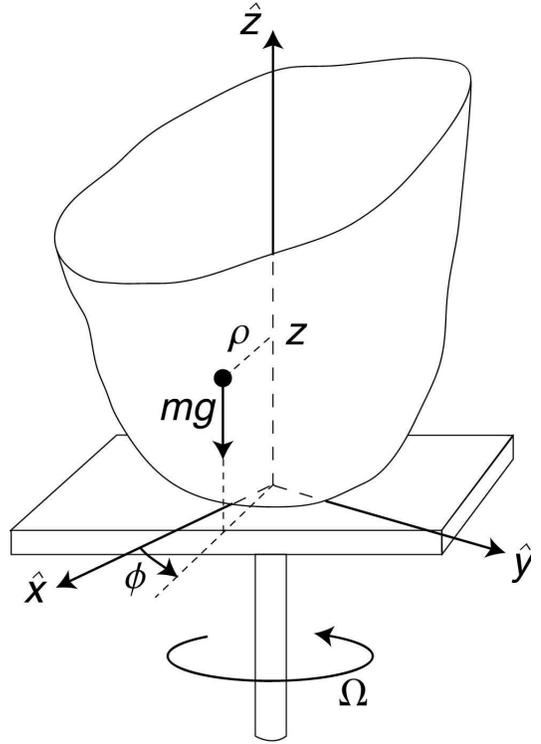,width=0.8\hsize}}
\smallskip
\caption{Particle of mass $m$ moving under gravity 
on the general surface $f=0$ of a bowl rotating at uniform speed $\Om$.}
\label{fig1}
\end{figure}

For suitable $f$,
this characteristic matrix is sufficiently general to 
model all four parameters $\Im A$, $\Re B_1$, $\Re B_2$, $\Re B_3$. 
However,
to generate a finite $\Re B_1$ in the absence of dissipation,
either $f_{\rh\ph}$ or $f_{\ph z}$ must be nonzero.

A special case,
used in the next section,
is a bowl with horizontal cross sections
near the equilibrium point
forming ellipses of constant eccentricity $e$
and semi-major axes with the same orientation.
For definiteness,
we consider the surface determined near the equilibrium point by
\beq
f(\rh,\ph,z) 
= \rh^2 (1 - e^2 \cos^2 \ph) - (1-e^2) (z/k)^{2/n}
=0.
\label{ellipsoid}
\eeq
This describes a bowl of uniform
elliptical horizontal cross section
and vertical cross section along the $x$ axis 
determined by $z=kx^n$. 
One of the two equilibrium points is at 
$\rh_0=(\Om^2/nkg)^{1/(n-2)}$,
$\ph_0 = 0$,
$z_0 = k \rh_0^n$.
Small oscillations about this point are described
in the $\rh$ and $\rh_0\ph$ coordinates 
by the characteristic matrix
\beq
X = \left(
\begin{array}{lr}
- (1 + \Ga^2)\om^2 + (n-2) \Om^2 
& -2i\Om\om 
\\ & \\
+2i \Om\om 
& - \om^2 + \ep^2 \Om^2
\end{array}
\right).
\label{xellipsoid}
\eeq
The component $X_{22}$ involves a function $\ep^2 (\ph_0)$,
which for later purposes is defined generally as
$\ep^2(\ph_0) = e^2(2 \cos^2\ph_0  - 1)/(1 - e^2 \cos^2 \ph_0)$.
In the present case $\ph_0 = 0$,
which gives $\ep^2 = e^2/(1-e^2)$.
We note in passing that the oscillatory motions 
determined by $X$ are stable for $n>2$.

In this simple model,
the term $\Re B_1$ vanishes.
The analysis in the next section shows that this 
can be avoided with the addition of appropriate dissipative terms.
However,
we note in passing that a nonzero $\Re B_1$
can be obtained without dissipation
by a relatively simple modification of the surface,
involving a helical twist with height.
The idea is to arrange matters so that 
the semi-major axis of the horizontal elliptical cross section
rotates as $z$ increases.
It suffices to replace $\ph$ in the bowl surface function $f$
of Eq.\ \rf{ellipsoid}
with a function $\ph+\th(z)$.
The equilibrium condition for $\ph$ becomes 
$\ph_0 = - \th(z_0)$,
and the characteristic matrix for small oscillations
acquires an additional contribution.
For example,
choosing $\th(z) = \ta z/\rh_0$ with constant $\ta$
produces a characteristic matrix equal to  
the sum of Eq.\ \rf{xellipsoid}
and a twist term $X_\ta$,
given by
\beq
X_\ta = \ep^2 \ta\Ga\Om^2 \left(
\begin{array}{lr}
\ta\Ga 
& 1
\\ & \\
1
& 0 
\end{array}
\right).
\label{xtwist}
\eeq
There is therefore a contribution to $\Re B_1$
determined by the twist constant $\ta$.

\section{Models with viscous damping}

In this section,
we consider analogue models involving classical oscillators 
with dissipation.
Since we are using linear and homogeneous equations of motion
and the correponding characteristic matrix
to model the neutral-meson effective hamiltonians,
we limit attention only to damping forces linear and homogeneous in the 
generalized coordinates and velocities.
We refer to such damping forces as viscous,
although this is a somewhat broader definition
than normally used by physicists.
Note that dry friction can also give linear equations of motion,
but typically leads to inhomogeneous terms
and so is disregarded here.  

The standard procedure in classical mechanics
is to obtain viscous damping forces 
from a Rayleigh dissipation function,
which is a symmetric quadratic form in the generalized velocities.
However,
under special circumstances viscous damping 
can lead to linear homogeneous damping forces involving
also the generalized coordinates
\cite{mr}.
This case is of direct interest in the present context.
We therefore work here with a generalized dissipation function $\cF$
that can handle damping in a broader class of models
\cite{lm}.

Up to irrelevant terms, 
we take $\cF$ to be a general quadratic expression
in the small-oscillation variables $Q$ and $\dot Q$:
\beq
\cF = \half \dot Q^T R \dot Q
+ \dot Q ^T H Q.
\label{diss}
\eeq
As usual, 
the damping forces are determined by the derivative of $\cF$
with respect to the generalized velocities.
The real symmetric matrix $R$ contains the standard Rayleigh 
dissipation matrix for viscous damping,
along with any contributions from other types of damping
that generate forces linear in the generalized velocities.
The real antisymmetric matrix $H$ determines damping forces 
linear in the generalized coordinates.

Combined with the Euler-Lagrange equations \rf{elnodamp},
the generalized dissipation function \rf{diss} 
leads to equations of motion for small oscillations
in the classical model given by
\beq
T \ddot Q + (G+R) \dot Q + (V+H) Q=0.
\label{eldamp}
\eeq
A harmonic solution has the form $Q=\Re[A\exp(i\om t)]$
as before,
but in the presence of damping $\om$ is complex.
We write
$\om = 2\pi \nu + i \ka= \mu + i \ka$.
In what follows,
we also use $\om^2 = \De^2 + 2 i \mu\ka$,
where $\De^2 = \mu^2 - \ka^2$.
To simplify the discussion,
we take the magnitude of the damping to be sufficiently small 
that potential complications 
such as the issue of stability 
require no special attention.

The characteristic matrix is
\beq
X = - T \om^2 + i (G+R) \om + V+H.
\label{xtgrvh}
\eeq
In terms of the parametrization in Eq.\ \rf{xparm},
we find
\bea
\Re A &=& 
\mu\ka (T_{11} + T_{22}) - \half \mu (R_{11}+ R_{22}),
\nonumber\\
\Im A &=& 
-\half \De^2 (T_{11} + T_{22}) 
\nonumber\\
&&\qquad
- \half\ka (R_{11}+ R_{22}) + \half (V_{11}+ V_{22}),
\nonumber\\
\Re B_1 &=& 
- \De^2 T_{12} - \ka R_{12} + V_{12},
\nonumber\\
\Im B_1 &=& 
- 2\mu\ka T_{12} +\mu R_{12},
\nonumber\\
\Re B_2 &=& 
- \mu G_{12},
\nonumber\\
\Im B_2 &=& 
- \ka G_{12} + H_{12},
\nonumber\\
\Re B_3 &=& 
-\half \De^2 (T_{11} - T_{22}) 
\nonumber\\
&&\qquad
- \half \ka (R_{11}- R_{22}) + \half (V_{11}- V_{22}),
\nonumber\\
\Im B_3 &=& 
-\mu\ka (T_{11} - T_{22}) + \half \mu (R_{11}- R_{22}).
\label{xdamp}
\eea
Inspection of these expressions
shows that a sufficiently general model
can indeed emulate independently
all eight real parameters in the effective hamiltonian
for a neutral-meson system. 
Note that the parameter $\Re B_2$ is unaffected by dissipation,
as mentioned in section III,
implying that a complete emulation of the neutral-meson system
\it requires \rm a nonzero gyroscopic term 
and therefore is most readily accomplished 
using a single two-dimensional oscillator. 
Note also that the damping force involving the matrix $H$ 
contributes only to $\Im B_2$,
whereas the matrix $R$ affects all parameters other than $B_2$.

As an explicit realization of these ideas,
we revisit the analogue model 
considered in section IV
describing a particle moving in a uniformly rotating bowl
with surface function $f$.
We suppose that the particle experiences 
an external viscous damping force.
This might be implemented with a mesh bowl
that allows resistance to the particle motion 
from the air or from some other static fluid 
in which the bowl and particle are immersed.
We take the generalized dissipation function for this resistance
to be
\beq
\cF = \half m h [\dot \rh^2 + \rh^2 (\dot\ph + \Om)^2 + \dot z^2],
\label{external}
\eeq
where the functional form of 
$\dot z = \dot z (\rh, \dot \rh,\ph,\dot\ph )$ 
is understood to be determined from the bowl surface equation $f=0$. 
We also suppose that the parameter $h$,
which controls the magnitude of the damping forces,
is sufficiently small to avoid difficulties with stability.

Inspection of the forces obtained from Eq.\ \rf{external}
reveals that an additional constant damping force $h\rh_0^2 \Om$ 
in the $\ph$ direction acts on the particle at equilibrium
and moves the equilibrium position away from the 
previously determined location. 
For example,
in the special case of a uniform elliptical horizontal cross section,
the equilibrium point is displaced from the apex of the ellipse.
In general,
the location of the new equilibrium point 
is determined by the simultaneous solution of the three equations 
$f = 0$,
$f_\rh + \Ga f_z = 0$,
and $f_\ph - \Si \rh_0 f_z = 0$,
where $\Si = h\Ga/\Om$.

The dissipation function $\cF$ in Eq.\ \rf{external}
describes the fluid resistance to the particle motion.
For small oscillations,
it includes both 
Rayleigh-type dissipation via a matrix $R$
and damping linear in $Q$ described by a matrix $H$.
It thus implements the form of Eq.\ \rf{diss}.
The associated equations of motion can be derived,
along with the accompanying characteristic matrix.
We find that the components of the characteristic matrix are 
the sum of the corresponding components in Eq.\ \rf{xcomps}
with additional terms given by the components of 
a matrix $\De X$:
\bea
\De X_{11} &=& 
ih (1 + \Ga^2)\om ,
\nonumber\\
\De X_{12} &=& 
\Ga\Si\om^2 +\la_0 \Si(f_{\rh z} + \Ga f_{zz}) 
- ih\Ga\Si\om ,
\nonumber\\
\De X_{21} &=& \De X_{12} +2 h\Om ,
\nonumber\\
\De X_{22} &=& 
-\Si^2 \om^2 + 2\la_0\Si f_{\ph z}/\rh_0 -\la_0 \Si^2 f_{zz}
\nonumber\\
&&\qquad
+ih(1 + \Si^2)\om .
\label{dex}
\eea
This result shows that 
the introduction of a relatively simple viscous damping force
suffices to ensure that all four parameters
$\Re A$, $\Im B_1$, $\Im B_2$, $\Im B_3$
can become nonzero.

For the special case of the bowl 
with uniform elliptical horizontal cross section 
described by Eq.\ \rf{ellipsoid},
the incorporation of viscous damping via Eq.\ \rf{external}
results in an equilibrium point at 
$z_0 = \Om^2\rh_0^2/ng$,
with 
$\rh_0^2 = [(1-e^2)^n\Om^4/(1-e^2\cos^2\ph_0)^n(nkg)^2]^{1/(n-2)}$
and 
$\tan\ph_0 = -(1 - \sqrt {1 - x})/a$,
where
$x = (1 - e^2)a^2$ and $a = 2h/e^2\Om$.
The requirement of real $\ph_0$ constrains the magnitude of $h$ 
to $|h|\leq e^2|\Om |/2\sqrt{1 - e^2}$.
The corresponding characteristic matrix for 
small oscillations in the $\rh$ and $\rh_0\ph$ coordinates
is given by the sum of Eq.\ \rf{xellipsoid}
with an additional matrix $\De X$.
In analogy with Eq.\ \rf{xtgrvh},
$\De X$ can be taken to have the form
\beq
\De X = - \De T\om^2 + i \De R \om + \De V+\De H.
\label{dextgrvh}
\eeq
Note that a putative term of the form $\De G$ is absent,
as expected.
The matrices $\De T$, $\De R$, $\De V$, $\De H$ are:
\beq
\De T = 
\left(
\begin{array}{lr}
0
&
-\Ga\Si
\\ & \\
-\Ga\Si 
\quad
& 
\Si^2 
\end{array}
\right),
\label{det}
\eeq
\beq
\De R = 
\left(
\begin{array}{lr}
h(1 + \Ga^2)
&
-h\Ga\Si
\\ & \\
-h\Ga\Si 
& 
h(1 + \Si^2)
\end{array}
\right),
\label{der}
\eeq
\beq
\De V = 
\left(
\begin{array}{lr}
0
&
- (n-1) h \Om
\\ & \\
- (n-1) h \Om 
\quad
& 
(n-2) h^2 
\end{array}
\right),
\label{dev}
\eeq
\beq
\De H = 
\left(
\begin{array}{lr}
0
&
 - h \Om
\\ & \\
h \Om 
\quad
& 
0
\end{array}
\right).
\label{deh}
\eeq

For small $h$ and hence small $\Si$,
the diagonal elements of the matrices $\De T$, $\De V$ 
can be viewed as perturbations on the 
result \rf{xellipsoid},
which involves nonzero $T$, $G$, and $V$.
However,
the contributions from 
the off-diagonal elements of $\De T$, $\De V$ 
and from $\De R$, $\De H$ 
are crucial for the complete emulation 
of a neutral-meson effective hamiltonian.
In particular,
Eq.\ \rf{xdamp}
shows that the parameters
$\Re A$, $\Im B_1$, $\Im B_2$, $\Im B_3$
are all nonzero,
as desired.

\section{Summary and discussion}

This paper studied the emulation 
of indirect CP violation in neutral-meson systems
using oscillator models in classical mechanics.
We obtained some no-go results for analogue models
without damping and for ones involving two one-dimensional oscillators. 
The implementation of spontaneous symmetry breaking 
was shown to be feasible. 
We proved that analogue models involving one two-dimensional oscillator
with rheonomic constraints
can suffice to emulate all eight real parameters
in the meson effective hamiltonian,
including the three describing physical T and CPT violation.

We presented a specific analogue model that provides a complete emulation.
It involves a particle moving under gravity 
on the surface of a uniformly rotating bowl of elliptical cross section
in the presence of weak external viscous damping.
The equations for small oscillations about an equilibrium point
are determined by a characteristic matrix 
given as the sum of $X$ in Eq.\ \rf{xellipsoid}
and $\De X$ in Eq.\ \rf{dextgrvh}.
The parametrization \rf{xparm} of this characteristic matrix,
with parameters fixed by Eq.\ \rf{xdamp},
can be placed in one-to-one
correspondence with the parametrization \rf{laparm}
in the $P^0$-$\overline{P^0}$ basis 
of the effective hamiltonian $\La$ for a neutral-meson system.
The correspondence is 
$A \leftrightarrow D$,
$B_1 \leftrightarrow E_1$,
$B_2 \leftrightarrow E_2$,
$B_3 \leftrightarrow E_3$.
Correspondences also exist with the effective hamiltonian
\rf{tilaparm} in the CP-eigenstate basis
or, since the emulation is complete,
in any other basis.

The results we have obtained leave open
some interesting issues. 
One is the extent to which 
quantitative values of experimental observables 
in any neutral-meson system
can be emulated in a realistic version
of the models we have discussed.
A satisfactory match would require reproducing the
relative sizes of the values 
of the masses, lifetimes, and parameters for CP violation.
Since the experimental data available on 
oscillations in the four neutral-meson systems
range from being relatively complete for the $K$
to limited for the $B_s$,
the degree of difficulty in obtaining a satisfactory match
varies considerably.
In any case,
the full flexibility of the analogue models is unnecessary because
no CPT violation has been observed to date
\cite{kexpt,dexpt,bexpt}.
For the special case of the $K$ system
CP violation is observed to be small,
so an emulation involving small dissipation 
is likely to be possible.
It would be interesting to determine the 
feasibility of constructing a quantitatively accurate model,
including perhaps constructing a working prototype.

A more ambitious task would be to explore the insights 
provided by the model about T and possible CPT violation 
with an eye to understanding its origin in nature
and the neutral-meson systems.
For example,
it is intriguing that the no-go results strongly favor
two-dimensional systems with rheonomic constraints.
This suggests a preference for a dynamical origin
of CP violation.
Similarly,
it is interesting that
the CPT violation in the analogue model 
emerges from a violation of rotation invariance.
This would appear to correspond 
with the situation in conventional quantum field theory
in the context of the standard model,
where the known mechanism for CPT violation
originates in the spontaneous violation 
of Lorentz symmetry
\cite{ks}
and implies CPT signals in neutral-meson systems
that depend on the orientation and magnitude
of the meson momentum
\cite{ak}.

Another interesting topic,
raised by Rosner 
\cite{jr},
is the emulation of T and CPT violation
by electrical circuits.
The general analysis we have provided in this work
can offer some insights.
A detailed analysis of this subject lies beyond the 
scope of this work,
but in what follows we provide a few remarks.

Suppose
that each of the strong-interaction eigenstates
$P^0$ and $\overline{P^0}$
is modeled as an oscillating electric circuit,
with CP violation regarded as a weak coupling between them.
For definiteness,
we view the two meson wave functions
as corresponding to the charges $q_1(t)$, $q_2(t)$
flowing through the circuits
as a function of time.
As in the case of the analogue model in classical mechanics,
the energies of the meson eigenstates are emulated 
by the oscillator frequencies
while the meson decay rates correspond to the oscillator dampings. 
Inductances in the circuit replace masses in the mechanical model,
inverse capacitances replace coupling constants in the potential,
and resistances provide dissipation.

The two-component meson wave function $\Ps$
can be identified with a two-component object $Q(t)$
formed from $q_1$ and $q_2$, 
as for the case of a classical-mechanics model.
We take the differential equations describing 
oscillations of the charges
in the circuit to be linear in $Q$ 
and its time derivatives $\dot Q = I$ and $\ddot Q = \dot I$,
where $I$ is the two-component current. 
In the absence of dissipation,
the equation governing the oscillatory behavior of $Q$ is 
Eq.\ \rf{elnodamp},
where the matrices $T$, $G$, $V$ are interpreted as 
characterizing appropriate properties of the circuit.
This means that much of the analysis 
in section III applies in modified form.
In particular,
the results obtained there reveal that a primary obstacle 
to a complete emulation of CP violation via electric circuits
is the need for an antisymmetric matrix $G$
coupling the currents in the two circuits.

A suitable circuit realization of $G$ requires a two-port device that 
is passive (no energy storage, increase, or dissipation).
The antisymmetry implies that reciprocity is broken:
a potential $\cV$ applied across the first port would induce 
a current across the second differing in phase by $\pi$ 
relative to the current induced across the first port when
the same $\cV$ is applied across the second.
Remarkably,
two-port devices of this type,
called gyrators,
have been the subject of some attention 
in the specialized electronics literature 
since their original invention by Tellegen in 1948 
\cite{bdht}.
Moreover,
a variety of network realizations of a gyrator exist
\cite{mg}.

We therefore suggest it is feasible to develop an electric circuit
emulating all eight parameters in the effective hamiltonian
for a neutral-meson system,
including both T and CPT violation.
A gyrator would implement the crucial T-violating features 
of Eq.\ \rf{elnodamp}
and in particular a nonzero $\Re B_2$ in Eq.\ \rf{xnodamp}.
As in the case of the classical-mechanics oscillators,
it would also be necessary to include dissipation.
This requires designing a circuit that incorporates
suitable damping elements 
leading to the oscillatory behavior given by Eq.\ \rf{eldamp}.
A general matrix $R$ can be obtained  
by a suitable placement of resistors in the circuit,
though producing a dissipation matrix of the form of $H$ 
might be less straightforward.
Developing an electrical 
realization of spontaneous symmetry breaking
would also be attractive.
A circuit designed to exhibit all these features
would make an impressive tabletop demonstration 
emulating T and CPT violation in neutral-meson systems.

The results obtained in the present work may also have application
in the emulation of other quantum oscillations in physics.
For example,
it would be of interest to study analogue models for 
neutrino oscillations.
A complete analysis for this case is likely to be more involved,
partly because three neutrino species are known
and the options for CP and CPT violation are correspondingly
more complicated.
Nonetheless,
an explicit analogue model in classical mechanics or with 
electric circuits could provide valuable insight.

\section*{Acknowledgments}

This work was supported in part 
by the United States Department of Energy 
under grant number DE-FG02-91ER40661.

\end{multicols}

\begin{references}
\bibitem{fn1}
We limit attention in this work to indirect CP violation.
\end{references}
\end{document}